\documentclass[aps,prl,twocolumn,superscriptaddress]{revtex4-2}
\usepackage{amssymb}
\usepackage{physics}
\usepackage{graphicx}
\usepackage{amsmath}
\usepackage{amsthm}
\usepackage{amsfonts}
\usepackage[T1]{fontenc}
\usepackage{array}
\usepackage{multirow}
\usepackage{color}
\usepackage{esint}
\usepackage{bm}
\usepackage{color}
\usepackage{bbm}
\usepackage{hyperref}
\usepackage{babel}
\usepackage{titlesec}

\begin{document}
	\title{Stable measurement-induced Floquet enriched topological order}
	
	\author{DinhDuy Vu}
	\affiliation{Condensed Matter Theory Center and Joint Quantum Institute, Department of Physics, University of Maryland, College Park, MD 20742, USA}
    \affiliation{Kavli Institute for Theoretical Physics, University  of  California,  Santa  Barbara,  CA  93106,  USA}

    \author{Ali  Lavasani}
    \affiliation{Kavli Institute for Theoretical Physics, University  of  California,  Santa  Barbara,  CA  93106,  USA}

    \author{Jong Yeon Lee}
    \affiliation{Kavli Institute for Theoretical Physics, University  of  California,  Santa  Barbara,  CA  93106,  USA}

    \author{Matthew P. A. Fisher}
    \affiliation{Department of Physics, University of California, Santa Barbara, CA 93106, USA}
 
\begin{abstract}
The Floquet code utilizes a periodic sequence of two-qubit measurements to realize the topological order. After each measurement round, the instantaneous stabilizer group can be mapped to a honeycomb toric code, explaining the topological feature. The code also possesses a time-crystal order – the $e-m$ transmutation after every cycle, breaking the Floquet symmetry of the measurement schedule. This behavior is distinct from the stationary topological order realized in either random circuits or time-independent Hamiltonian. Therefore, the resultant phase belongs to the overlap between the classes of Floquet enriched topological orders and measurement-induced phases. In this work, we construct a continuous path interpolating between the Floquet and toric codes, focusing on the transition between the time-crystal and stationary topological phases. We show that this transition is characterized by a divergent length scale. We also add single-qubit perturbations to the model and obtain a richer two-dimensional parametric phase diagram of the Floquet code, showing the stability of the Floquet enriched topological order.
	\end{abstract}
	
	\maketitle

 \textit{Introduction - }Topological phases are highly interesting because their non-local integrals of motion are robust against local perturbations and thus beneficial for fault-tolerant quantum computation \cite{Dennis2002,Kitaev2003,Kitaev2006,Levin2005,Hastings2005}. These exotic features were first established for the ground states of certain time-independent Hamiltonians such as $\mathbb{Z}_2$ toric code, and subsequently were extended to non-equilibrium unitary dynamics \cite{Rudner2013,Po2016,Potter2016,Else2016b,Rudner2020,Vu2022a} as well as non-unitary dynamics that involve measurements \cite{Li2018,Skinner2019,Chan2019,Bao2020a,Zabalo2020,Chen2020,Lavasani2020,Bao2021a,Lavasani2021,Wampler2021,Lavasani2022,Hastings2021,Ippoliti2021,Sriram2022}. 
 In particular, Hasting and Haah proposed a circuit model consisting of a sequence of two-body measurements that gives rise to a dynamically generated quantum error correcting code which is closely related to $\mathbb{Z}_2$ toric code. Due to the time-periodic nature of the measurement sequence, the underlying code and logical operators  transform periodically in time; thus the protocol code is called the Floquet code \cite{Hastings2021}
This code is not a stabilizer nor subsystem code, but dynamically generates logical qubits through a sequence of non-commuting measurements \cite{Hastings2021,Aasen2022,Kesselring2022,Davydova2022}.

A remarkable characteristic of this class of Floquet codes is the time-crystalline ordering that transmutes magnetic ($m$) to electric ($e$)-type logical operators and vice versa each cycle.
Unlike conventional time crystals that can be accessed by local order parameters \cite{Else2016,Khemani2016,von2016}, the $e-m$ exchange can only be accessed from non-local operators,
suggesting its topological nature. In fact, on an open boundary, this non-trivial exchange manifests as a radical chiral mode with a dynamical topological invariant $\sqrt{2}$ \cite{Po2017,Potter2017}, and is thus dubbed as the Floquet enriched topological order (FET). The focus of this work is a measurement-induced version of FET.
 
 \begin{figure}
	\centering
	\includegraphics[width=0.45\textwidth]{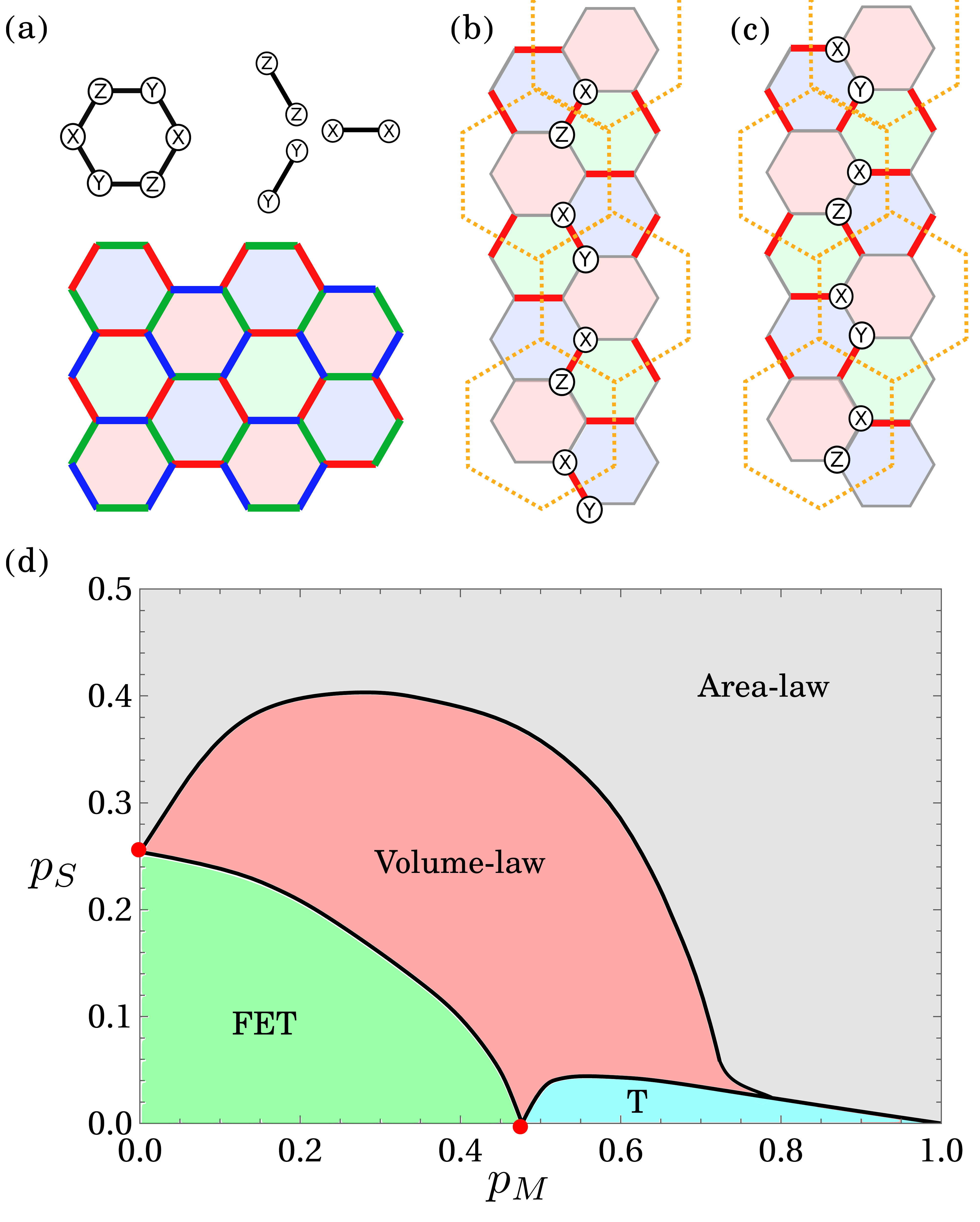}
	\caption{(a) Layout of the Floquet code with color labels for each link and plaquette and the explicit forms of the plaquette and link operators. The link operators only depend on the respective link orientation and not on the color. (b-c) One configuration of the $m$ and $e$ strings defined at the red round, respectively.
	(d) Phase diagram with respect to the missing $p_M$ and single-qubit measurements $p_S$ probabilities.}
	\label{phasediagram}
 \end{figure} 

The main goal of this work is to study the stability of the Floquet topological order against perturbations of the circuit model away from the ideal protocol in Ref.\cite{Hastings2021}. Although some ideas of fault tolerance have been discussed in\cite{Hastings2021}, our work illuminates two aspects. First, it establishes the measurement-induced FET as a phase of matter, spanning a finite region of the parameter space. Second, for quantum platforms with native measurement operations, measurement-induced automorphisms of logical operators can be used as logical gates for quantum computing, e.g. the $e-m$ exchange facilitated by the Floquet code is equivalent to a Hadamard gate that rotates between logical $X$ and $Z$. Our work shows that such logical gates inherit the Floquet topological nature and are thus fault-tolerant. For demonstration, we consider the effect of skipping some of the measurements as well as the effect of randomly replacing a two-qubit measurement with a pair of single-qubit measurements. Interestingly, by using an \textit{error-corrected order parameter}, we find that the Floquet order is robust against small but finite perturbations, giving rise to an extended measurement-induced Floquet topological phase of matter. As one increases the perturbation strength, there is a measurement-induced phase transition to either a non-Floquet topological phase, a volume law phase, or a trivial area law phase (Fig. \ref{phasediagram}). We note that the FET phase, being stabilized  by measurements, does not undergo thermalization in the long time limit. This is not the case for unitary models without either MBL \cite{Lazarides2015,Abanin2016,Po2016,Potter2016,Else2016b} or prethermalization \cite{Else2017,Vu2022,Hart2022}.
 
  \begin{figure*}
	\centering
	\includegraphics[width=0.95\textwidth]{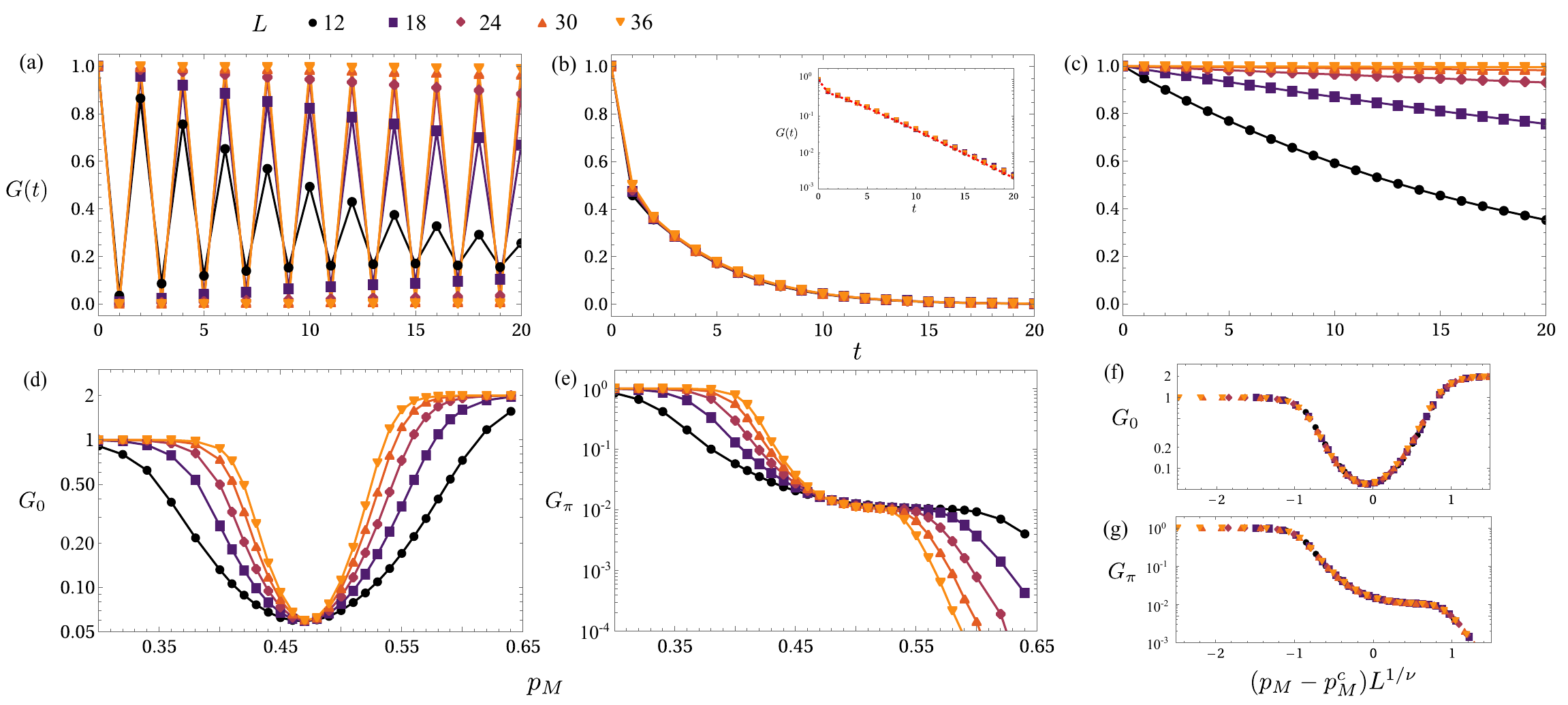}
	\caption{(a-c) $G(t)$ for $p_M=0.38$ (FET), $0.48$ (critical), and $0.58$ (T). The FET phase features double-period oscillations, which is lacking in the T phase. The inset of (b) shows $G(t)$ at the critical point fitted to a multi-channel Markovian process. (d-e) The $0$ and $\pi$-components averaged over $T=100$ cycles. (f-g)  Data collapse of (d-e) respectively with $(p_M^c,\nu)=[0.47(8),1.36(0)]$ (f) and $[0.47(6),1.35(4)]$ (g).}
	\label{G_t}
\end{figure*} 
 
\textit{Model - } 
In this study, we consider a periodic-boundary $L\times L$ honeycomb lattice where a qubit is located at each vertex, with $2L^2$ qubits in total. Each hexagonal plaquette is assigned one of three colors: red, blue, or green, ensuring that no two adjacent plaquettes share the same color.
Within this geometry, a link only connects plaquettes sharing the same color and thus inherits the respective color label. Furthermore, a two-qubit operator is assigned to each link, depending on the link orientation as demonstrated in Fig.~\ref{phasediagram}(a). A plaquette operator is defined as the clockwise product of the link operators around the corresponding plaquette. For the standard Floquet code, all plaquette operators are going to have a definite value (either $+1$ or $-1$), and since they commute with all link operators, their value will not change in the subsequent steps. 

Each link measurement projects a $\mathbb{C}^4$ Hilbert space of two adjacent physical qubits to a $\mathbb{C}^2$ Hilbert space which can then be regarded as a single effective qubit sitting on the link of a superlattice. Together with all the measured plaquette operators, these stabilizers make up a toric code state on a hexagonal supperlattice \cite{Hastings2021} (See~Fig.\ref{phasediagram}). 
Note that an $f$ logical operator is simply given by the product of all link operators along a non-trivial loop in the original lattice. 

We consider two types of perturbations to the standard Floquet code. First we consider randomly missing link measurements at the blue and green rounds with probability $p_M<1$, while the red round is free of missing defects. We refer to this perturbation as the ``missing'' ($M$-) perturbation. Additionally, given that a link is measured, with probability $p_S$ we replace the two-qubit measurement in the standard protocol with two disjoint single-qubit measurements, e.g, instead of measuring $X_1X_2$ on a link, we measure $X_1$ on one end and $X_2$ on the other end. We denote this perturbation as the ``single-qubit'' ($S$-) perturbation. This type of error is qualitative similar to random Pauli errors but simplified by the structure of the Floquet code \cite{Hastings2021}. $M$-perturbations commute with all plaquette operators, leaving the quantum state always topological at the end of each measurement cycle. Meanwhile, $S$-perturbations anti-commute with the plaquettes and can drive the quantum state to a different phase.

We choose the initial state of the circuit to be the common eigenstate of all the plaquette operators and the logical operators $m_x$, $m_z$; and study the entanglement structure of the late time state of the circuit after $\sim\mathcal{O}(L)$ evolution. We note that due to the random dynamics of the circuit, the late time entanglement structure would be independent of the choice of the initial state \cite{SM}. Nevertheless, our choice of initial state optimizes the readout of the time-crystalline order and reduces the time for the quantum state to reach the topological sectors.
 
 
To obtain the phase diagram, we compute the tripartite mutual information (TMI) of the late time state averaged over different random realizations, which equals one in the topological phase, zero in the trivial area-law phase, and a negative extensive value in the volume-law phase \cite{Gullans2020,Zabalo2020,Lavasani2022,Ippoliti2021}. The exact phase boundary can be obtained from a standard finite-size scaling method which also agrees with the transition in purification dynamics \cite{Gullans2020,Gullan2020b}. The detailed methodology and TMI data are presented in SM \cite{SM}. 
The phase diagram is shown in Fig.~\ref{phasediagram}(d).
Under the absence of $S$-perturbations, i.e., $p_S=0$, the diagram features a tricritical point at $p_M^c \approx 0.48$ which separates the FET phase with a period-doubled $e-m$ exchange behavior from the stationary topological (T) phase for $p_M$ strictly less than unity.

\textit{Missing link measurements--} 
First, we examine the phase diagram along the line $p_S = 0$ (no $S$-perturbation).
Because we have assumed no red link measurement would be missed, the bulk ISG is always the same (up to $\pm 1 $ signs) after each round of red link measurements. This leaves room for only the non-local logical qubits to have non-trivial dynamics. The limit $p_M=0$ realizes a standard honeycomb Floquet code equipped with an $e-m$ exchange every cycle. In the opposite limit $p_M\to 1$, the other two color rounds are essentially absent and the circuit generates a stationary toric code topological order on the superlattice associated with the red links. 

As we will show, the $e-m$ exchange dynamics persists for $p_M>0$ up to a critical value $p_M^c>0$, after which it disappears and the system enters stationary toric code phase. To probe the Floquet dynamics, we consider 
the following procedure.
Let $m_x$ and $m_z$ denote the $m$-type string operators winding around the torus along x and z directions respectively. $e_x$ and $e_z$ are defined analogously to denote the logical $e$-type string operators.  We start with the state which is the eigenstate of all plaquette operators as well as $m_x$ and $m_z$. 
After every red round of measurements, we read out the expectation value of the $m$-loop along the x-direction $G(t)=\overline{\langle{m_x(t)}\rangle^2}$, where 
the overline stands for averaging over random circuit realizations. For numerical simulation, we use Clifford formalism \cite{Aaronson2004}, where post-selection and averaging over measurement outcome are included. Note that for the initial state, we have $\expval{m_x(0)}=1$ and $\expval{e_x(0)}=0$. Therefore, due to the ${e-m}$ exchange dynamics, we expect $G(t)$ to oscillate between $1$ and $0$ in the FET phase, while it should remain constant and equal to $1$ in the toric code phase. 

By inspecting $G(t)$ with varying $p_M \in [0,1]$, we find an extended FET phase separated from the toric code topological phase by a sharp phase transition at $p_M^c=0.48$. More specifically, for $p_M < p_M^c$, the value of $G(t)$ in odd and even cycles are visibly distinguishable, and $G(t)$ converges to the exact binary form $G(t)=(t+1) \text{ mod } 2$ as $L\to \infty$, signaling the time-crystalline order (Fig.~\ref{G_t}a). On the other hand, for $p_M>p_M^c$, $G(t)$ approaches a single-valued function $G(t)=1$ in the thermodynamic limit (Fig.~\ref{G_t}c). Remarkably, at the critical point $p_M^c=0.48$ which separates the two phases, $G(t)$ acquires a finite lifetime and follows a universal function independent of the system size (Fig.~\ref{G_t}b), suggesting a zero dynamical exponent $z=0$. This unusual $z$ is due to the exact recovery of the bulk ISG after every cycle. Away from the $p_S=0$ axis, the dynamic exponent returns to the conventional value of unity \cite{SM}. 

 
Accordingly, one can use the $0$ and $\pi$  components of the Fourier transform of $G(t)$ as order parameters for distinguishing the two phases and performing scaling analysis:
 \begin{equation}
 	G_0 \equiv \lim_{T\to \infty}\frac{2}{T}\sum_{t=0}^T G(t),~G_\pi \equiv \lim_{T\to \infty}\frac{2}{T}\sum_{t=0}^T e^{i\pi t} G(t).
 \end{equation}
We note that the time limit should be taken after the thermodynamic limit, given that for any fixed system size, $\lim_{t\to \infty} G(t)=0$ [see Fig.~\ref{G_t}(a-c)].
Fig.~\ref{G_t}(d-e) shows the transition of these quantities across the critical point, taking $T=100$; we have $G_0=G_\pi = 1$ for $p_M<p_M^c$, while for $p_M>p_M^c$, we have $G_0=2$ and $G_\pi = 0$, in the thermodynamic limit. Both quantities follow the scaling form $F_{0,\pi}\qty[(p-p_M^c)L^{1/\nu}]$ near the criticality,
with $p_M^c\sim 0.48$ and $\nu\sim 1.35-1.36$ [see Figs.~\ref{G_t}(e-f)]. 

The nature of this phase transition can be understood by  the microscopic dynamics of the Floquet phase. As is explained in Ref.~\cite{Hastings2021}, the color-wise measurement of the link operators along a loop maps the $e$-string along that loop to an $m$-string along the same loop and vice versa (see also the SM~\cite{SM}). Therefore, whenever there exists a non-trivial path in the perturbed circuit model along which all link operators get measured in a cycle, the corresponding $e$-string operator along that path gets mapped to an $m$-string operator. However, due to the topological nature of the phase, it means \textit{any} $e$-string operator maps to the corresponding $m$-string operator, as long as there exists one path along which all link operators get measured properly. This picture places the FET-T transition in the same universality class as 2D percolation. More specifically, given that red links are always measured, 
one can contract each red links into a point 
and consider the bond percolation problem on the resulting Kagome lattice. In fact, the critical value $p_M^c=0.48$ for the FET-T phase transition agrees with the numerical estimate of bond percolation threshold on the Kagome lattice~\cite{Robert1997}, which explains the extended FET phase for $p_M<p_M^c$. On the other hand, when $p_M>p_M^c$, there is no percolating path, and hence each cycle consists of only disconnected finite-size patches of measurements. Since such measurements cannot access logical information, the $e-m$ exchange vanishes in the non-percolating phase.  The percolation picture also explains why $G_0$ vanishes at $p_M\,{=}\,p_M^c$(Fig.~\ref{G_t}d). At criticality, there is a finite chance of forming a percolating cluster whose boundary is also percolating around the torus. 
The result is a Markovian process that we show in the inset of Fig.~\ref{G_t}(b) and explain in more detail in the SM~\cite{SM}. 
We note that the Markovian nature is lifted if red rounds are imperfect so that errors persist from one cycle to the next and accumulate with time. This is the case of the transition along the $p_S$ axis in Fig.~\ref{phasediagram}(d). Another option is to introduce $M-$errors to red rounds. Both transitions are controlled by the 3D percolation universality \cite{SM}, signifying the extension along the time direction.

 \begin{figure}
 	\centering
 	\includegraphics[width=0.45\textwidth]{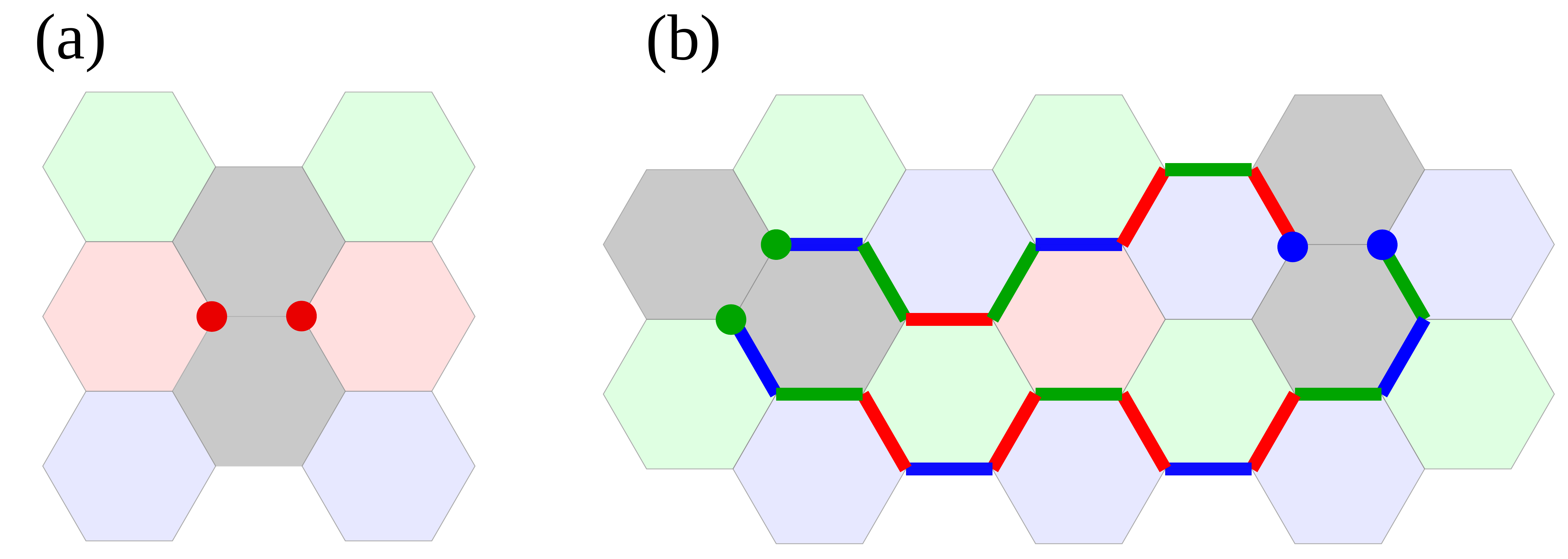}
 	\caption{Examples of errors that can occur to the ISG upon introducing $S-$perturbations. (a) A local red $S-$perturbation that can be detected through two adjacent plaquettes (gray). (b) Non-contractible open $f-$string formed when at least two green or blue $S-$perturbations intercept an $f-$loop.}
 	\label{openstring}
 \end{figure}
 
\textit{Single-qubit measurements - } 
Even though the Floquet code measures closed $f-$operators, these loops can be deformed into each other except for the two non-trivial loops winding around the torus and thus do not contribute to the entanglement inside the bulk. 
Therefore, the TMI stays exactly unity everywhere along the $p_S=0$ axis. The picture changes significantly when we introduce single-qubit perturbations. In fact, the generators of the ISG after each red measurement round now contain ``errors'' to the Hasting-Haah protocol not found in the $p_S=0$ case. In Fig.~\ref{openstring}(a), we show an error generated by pair of red single-qubit measurements that can be detected through the two neighboring plaquettes with which this error anti-commutes. Another type of errors occurs if more than one green or blue $S-$perturbations happens along an $f-$loop, resulting in open string errors that are not immediately transparent under plaquette measurements as shown in Fig.~\ref{openstring}(b). We note that at the tricritical point, the circuit measures diverging $f-$loops, which 
 under non-zero $p_S$, break into extended strings. This explains the immediate emergence of the volume phase at the critical point with any non-zero $p_S$ shown in Fig.~\ref{phasediagram}(d).    
 
We expect that the FET characteristic must survive throughout the topological phase with $p_M<p_M^c$ as shown in Fig.~\ref{phasediagram}(d) because there cannot be a continuous deformation between the FET and T phases. 
However, our previous probe cannot be applied straightforwardly in the presence of $S-$perturbation because the string that we read out may randomly cross a point or an error string. Even worse, the probability of crossing an error string increases with system size, making the defined order parameter to vanish in the large$-L$ limit. This is similar to the problem that the expectation values of loop operators decay exponentially with a perimeter-law when the toric code Hamiltonian is subjected to an Ising field \cite{Hastings2005}. Nevertheless, the topological order or the ground state degeneracy is preserved and the exact 1-form symmetry characterizing the $x-$loop is replaced by an emergent 1-form symmetry defined along a `fattened' loop with finite transverse width. 

\begin{figure}
  	\centering
 	\includegraphics[width=0.48\textwidth]{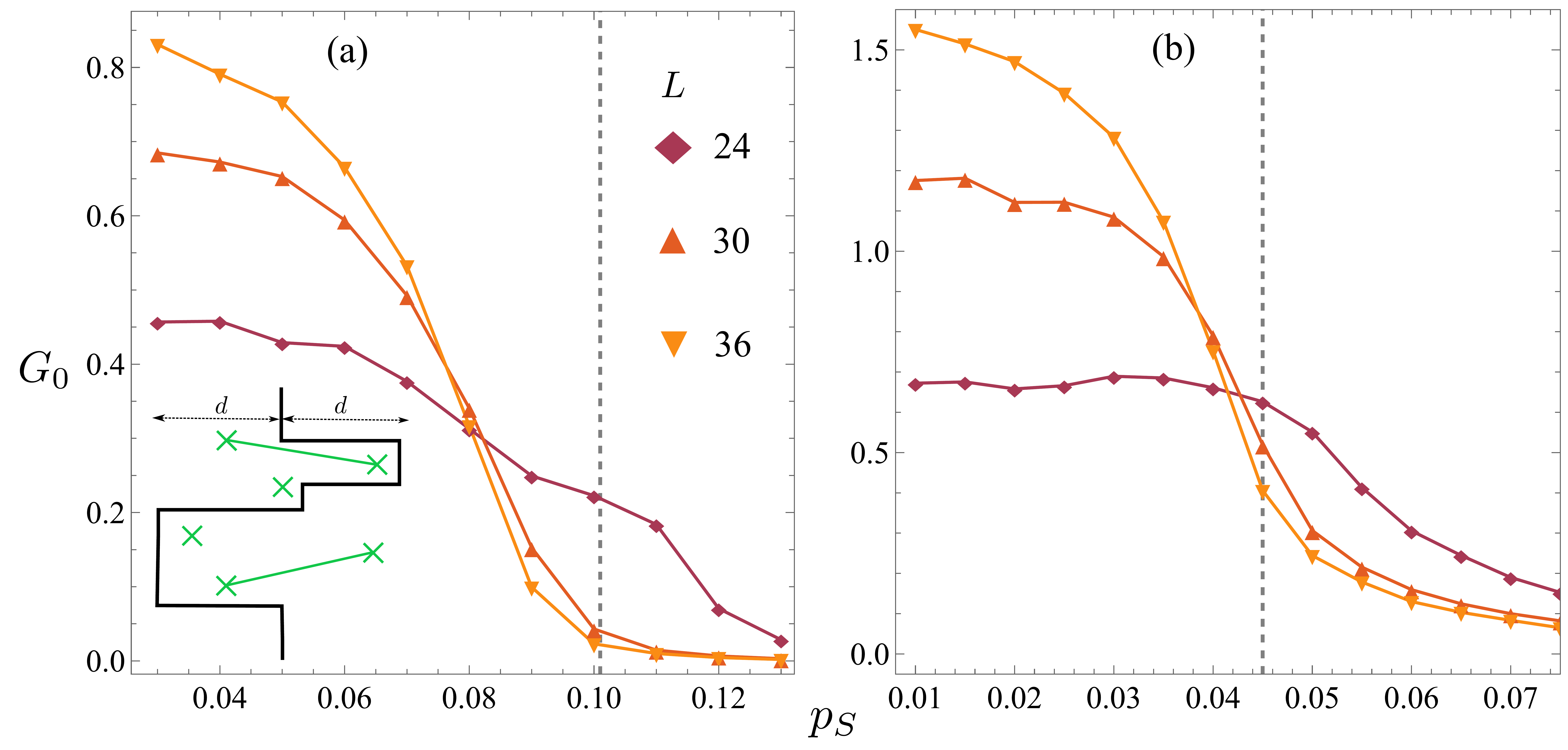}
 	\caption{The 0-component amplitude of $G$ across the FET-volume-law transition at $p_M=0.4$ (a) and the T-volume-law transition at $p_M=0.55$ (b). The dash lines indicate the actual critical $p_S$ obtained from TMI. The inset in (a) shows the corrected $m-$string after being morphed to avoid all errors.}
 	\label{pe}
 \end{figure}

The same intuition can be applied to our model. Specifically, by deforming our $m-$loop to avoid all the errors [see the inset of Fig.~\ref{pe}(a)], we can recover the finite-value order parameter as shown in Fig.~\ref{pe}. The exact implementation is described in the SM \cite{SM}. The typical width $d$ of the corrected $m-$string is governed by the length of string errors, which is finite in the topological phase but grows with the system size in the volume-law phase. In our correction scheme, to be compatible with the finite size, we fix the maximum width as $d=11$. Despite being independent of the system size in the area-law phase, $d$ should increase as the circuit approaches the phase transition, so this $d-$fixing makes the apparent transition in Fig.~\ref{pe} happen before the actual one computed from TMI; this is analogous to the situation where optimal and non-optimal decoders perform similarly away from the phase boundaries in the circuit with measurements~\cite{Lee2022}.
Nevertheless, this corrected order parameter justifies the robustness of the non-trivial $e-m$ exchange against single-qubit perturbations, which is expected for a topological phase.
 
 \textit{Conclusion - }We discover a robust Floquet topological phase in a quantum circuit characterized by a non-trivial exchange that brings the logical operator back to itself after two cycles.  We establish in our quantum circuit model that this phase is stable against microscopic perturbations to the protocol, thus verifying the topological nature of the phase. Additionally, even though the dynamics here is generated from measurements only, the phase diagram displays a rich structure with a volume-law phase \cite{Ippoliti2021}.
 
 The research on Floquet topological order can advance in several directions. An outstanding question is the bulk invariant that can capture FET. For non-interacting Floquet topology, the invariant describes a topological gap-closing point in the evolution unitary \cite{Rudner,Vu2022a}; by analogy, the invariant for FET presumably describes a non-abelian defect. Lastly, we note that our order parameter $G(t)$ is not linear in the density matrix, meaning that it can only be accessed experimentally with post-selection. 
 However, in principle, it should be possible to use the outcomes of the link measurements to perform proper error detection and error correction, similar to the standard Floquet code, to avoid this problem.  
 
\begin{acknowledgements}
\textit{Acknowledgements - }This research was supported in part by the National Science Foundation under Grant No. NSF PHY-1748958, the Heising-Simons Foundation and the Simons Foundation (216179, LB). The authors acknowledge the University of Maryland supercomputing resources made available for conducting the research reported in this paper. DV is supported by the Laboratory for Physical Science. JYL is supported by the Gordon and Betty Moore Foundation under the grant GBMF8690 and by the National Science Foundation under the grant PHY-1748958. M.P.A.F is supported by the Heising-Simons Foundation and the Simons Collaboration
on Ultra-Quantum Matter, which is a grant from the Simons Foundation (651457).
\end{acknowledgements}

\bibliographystyle{apsrev4-2}
\bibliography{reference}

\end{document}


\title{Supplemental Material for Stable measurement-induced Floquet enriched topological order}
	\maketitle
	
	\section{Phase diagram of the Floquet honeycomb code}
	In the ideal honeycomb Floquet code, two-qubit measurements are performed in cycles containing three rounds following the blue-green-red sequence. Each two-body operator defined on a link is called \emph{check} operator. After each round, the instantaneous stabilizer group (ISG) consists of all the plaquette operators (product of all the check operators on links around one hexagonal plaquette) and all the links measured in that round. For a honeycomb lattice of $L$ plaquettes per direction, there are $L^2$ plaquettes, $L^2$ links of one color, and $2L^2$ qubits. Not all of the stabilizers are independent because the product of all plaquettes is identity and so is the product of all links and plaquettes of the other two colors (e.g., the product of all the red links with all the blue and green plaquettes), resulting in two logical qubits. We fix these degrees of freedom by measuring the loop operators winding around the torus. 

\begin{figure}
		\centering
		\includegraphics[width=0.4\textwidth]{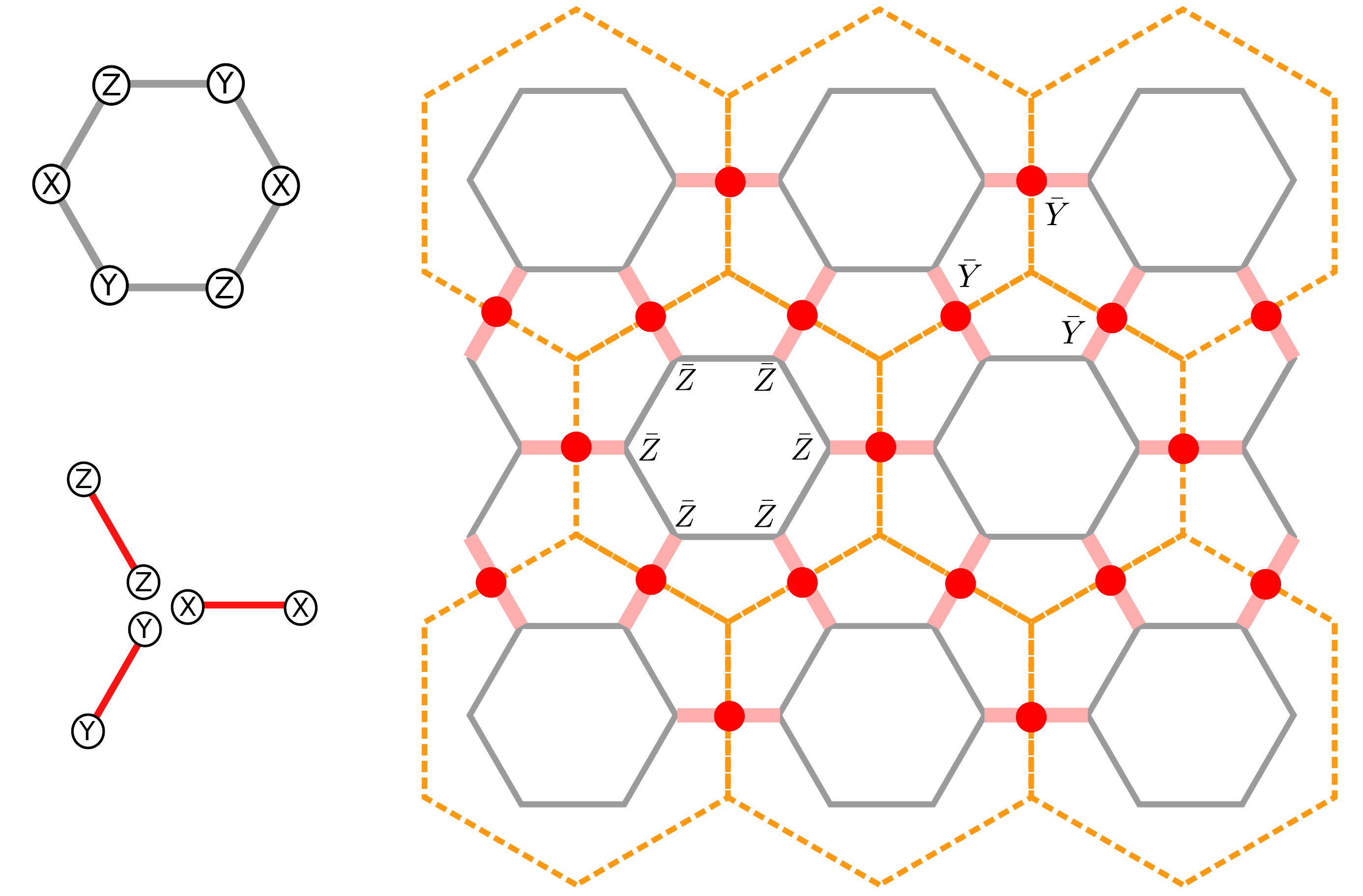}
		\caption{Left: Explicit expression of the physical plaquette and link operators. Right: By mapping a measured link (light red) into an effective qubit (deep red), the lattice is redrawn into a hexagonal superlattice (orange). The physical plaquette operators now serve as either the super-plaquette or super-vertex operators in the new lattice. \label{mapping} }
	\end{figure}

    For the standard Floquet code, the ISG can be mapped exactly to that of a toric code on a hexagonal lattice. Specifically, each link operator projects the 4D Hilbert space of two physical qubits to a 2D Hilbert space of one particular parity, which can be regarded as an effective qubit. For example, under a $ZZ=+1$ measurement the Hilbert space of this effective qubit in terms of the two physical qubits is $\{ \ket{\uparrow,\uparrow}\ket{\downarrow,\downarrow} \}$. Within this basis, we have the following transformations
    \begin{equation}
        \ket{\uparrow\uparrow} \xrightarrow{Z\otimes\mathbb{I}}  \ket{\uparrow\uparrow},~ \ket{\downarrow\downarrow} \xrightarrow{Z\otimes\mathbb{I}}  -\ket{\downarrow\downarrow};~
    \ket{\uparrow\uparrow} \xrightarrow{X\otimes Y}  i\ket{\downarrow\downarrow},~ 
    \ket{\downarrow\downarrow} \xrightarrow{X\otimes Y}  -i\ket{\uparrow\uparrow}. 
    \end{equation}
    As a result, considering the action on the effective qubit, we can regard $Z\otimes \mathbb{I} \equiv \tilde{Z}$ and $X\otimes Y \equiv \tilde{Y}$. In Fig.~\ref{mapping}, if we redraw the quantum circuit in effective qubits (red dots), the lattice is now transformed to a hexagonal superlattice. Additionally, a physical plaquette operator inside a superlattice cell is now rewritten as $\Pi \tilde{Z}_i$ with $i$ belonging to the super-plaquette; while a physical plaquette overlapping a superlattice vertex, in the new basis, is $\Pi \Tilde{Y}_j$ with $j$ surrounding the super-vertex. This set of effective plaquette and vertex operators in the superlattice generates the topological order despite the physical operators being only two-qubit. We can also see that the string operators in Fig.~1(b) and (c) of the main text, within the basis of effective qubits, become the $\Tilde{Y}-$ and $\Tilde{Z}-$strings respectively. 

	\begin{figure}
		\centering
		\includegraphics[width=0.9\textwidth]{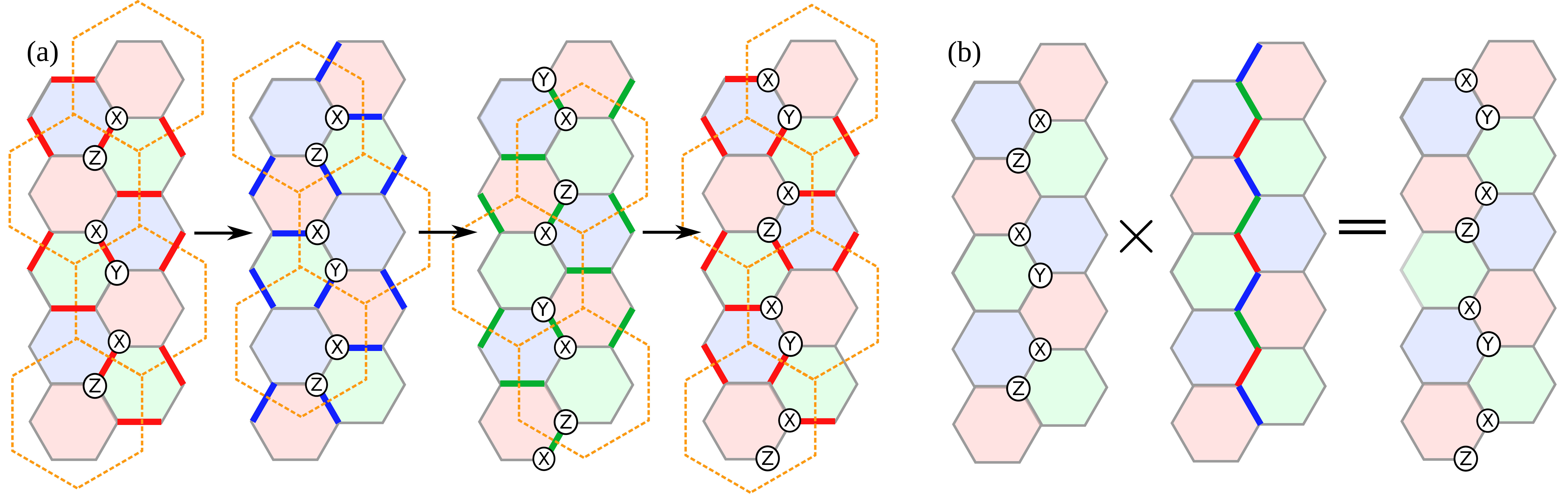}
		\caption{(a) Microscopic evolution that convert an $m$-type string to an $e$-type in the ideal Floquet code. (b) The equivalent picture of multiplying the $m-$type Pauli string with an overlaying $f-$string that produces the same $e-m$ transmutation. \label{Fig1}}
	\end{figure}
	
  \subsection{Floquet-to-stationary transition}	
  The Floquet code exhibits a time-crystalline behavior in which an $m$-type logical qubit is converted to an $e$-type after one cycle, breaking the time translational symmetry. The exchange can be figured out exactly in the ideal Floquet code where one sequentially updates the $m$-string with non-commuting measurements of the following blue, green, and red rounds as shown in Fig.~\ref{Fig1}(a). Operationally, this is equivalent to fusing the logical string operator with an $f$-string as shown in Fig.~\ref{Fig1}(b). In the ideal Floquet code where all links are measured, such an $f$-string is always possible. However, if only links of one or two colors are measured, we can construct at best closed plaquettes. This intuition suggests that we can look at the history of the measurement schedule to study the logical qubit exchange. 
  
\begin{figure}
  	\centering
  	\includegraphics[width=0.7\textwidth]{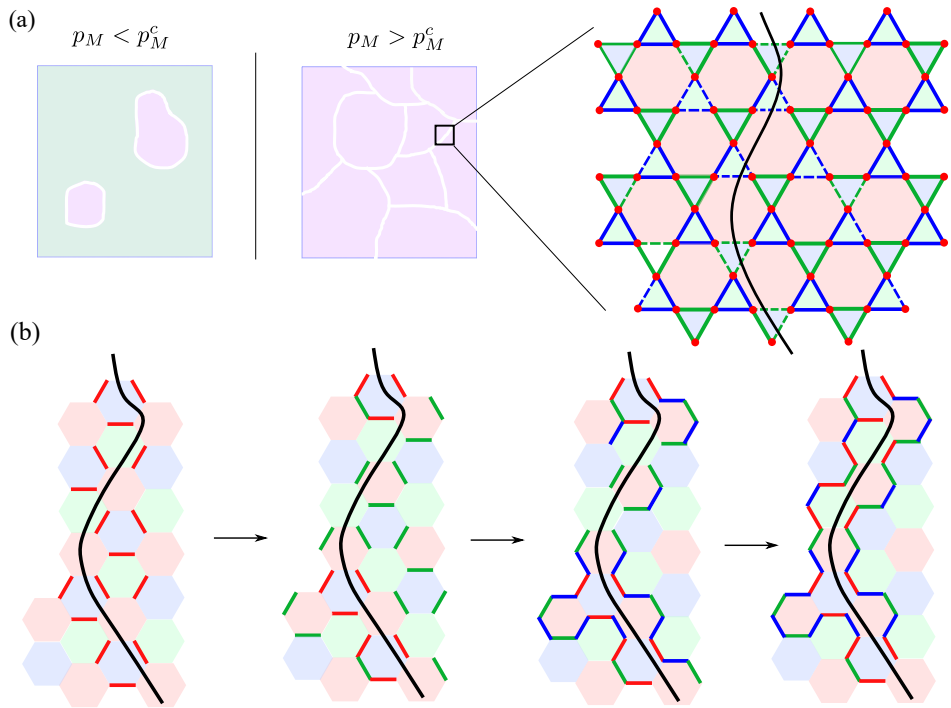}
  	\caption{(a) Collection of all measured links with ($p_M<p_M^c$) and without $(p_M>p_M^c)$ percolating clusters that extends throughout the system (green shaded). The record of the measurement schedules near the cluster boundary which runs through missing links (dashed lines) is shown on the right. The red links are contracted, resulting in a percolating kagome lattice. (b) The update of string operators in the ISG along the cluster boundary after the green, blue, and red measurement rounds. After the red round, an $f-$string is measured along the cluster boundary.}
  	\label{Fig3}
  \end{figure}

  Inspired by Ref.~\cite{Kesselring2022}, we look at the collection of all the links being measured in a cycle, which form a periodic 2D lattice if no links are missing. We define a cluster as a region inside which one can construct an $f$-path between any points. If not many links are missing, there exists a percolating cluster extending throughout the lattice; while on the opposite limit, all the clusters are finite and it is the cluster boundary that percolates through the lattice (Fig.~\ref{Fig3}a). This definition of the cluster exactly matches the FET-T transition to a percolation transition on a kagome lattice with each link being turned on randomly with probability $1-p_M$. In the percolating phase, one can always construct a path from a point to infinity, corresponding to our FET phase where the measurement history contains some extended $f-$string that implements the $e-m$ exchange. In contrast, in the non-percolating phase, such an $f-$string is not present, meaning the logical qubit does not transmute. This mapping yield the $p_M^c=1-p_\text{threshold}\approx 0.476$ and $\nu=4/3\approx 1.33$. Besides the transition from double to single-period oscillation of the logical qubit, we also observe a critical damping at the critical point, i.e., the lifetime of the logical qubit does not diverge with the system size. This phenomenon can also be inferred from the percolation picture. If the measurement operator simultaneously anti-commutes with multiple non-overlapping strings, the result is a concatenated longer Pauli string. This special condition is met along the boundary cluster. In Fig.~\ref{Fig3}(b), we show the growth of an $f-$loop along the cluster boundary.
  After the green and blue rounds, there exist open segments to the two sides of the cluster boundary which are not allowed in the perturbation-free Hasting and Haah code. Finally, at the red round, these segments are concatenated to form a continuous $f-$loop. At the critical point, the typical length of cluster boundaries diverges, resulting in a finite probability of measuring an infinitely long $f-$loop even in \textbf{one} cycle, leading to the critical damping. 
  
\begin{figure}
  	\centering
  	\includegraphics[width=0.97 \textwidth]{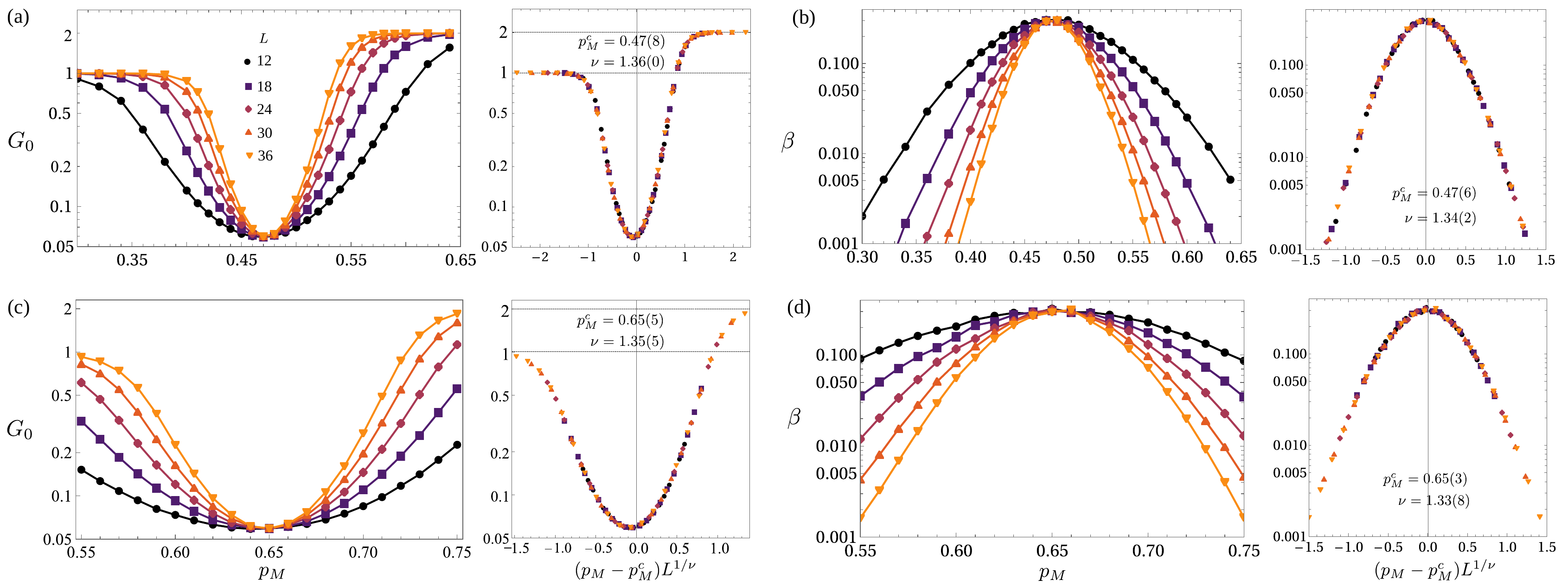}
  	\caption{(a) The zero component of the logical qubit oscillation averaged over 100 cycles. (b) Decay rate of the logical qubit over time. Data collapse of the respective figure is shown on the right. Both are for the original model with $M-$perturbations in the green and blue rounds. (c-d) Similar to (a-b) but the $M-$perturbations are only present in the green round. The phase transition is unchanged except for a shift of the critical $p_M$.}
  	\label{Fig2}
  \end{figure}

In principle one can find  the probability of code measuring an $f-$string at the critical point from the percolation picture but such a calculation would be quite involved. Instead, we obtain the relevant probabilities numerically from the evolution \textbf{over one cycle}. By initializing the circuit with two $m-$strings along $x$ and $z$ direction and looking at the probability of different outcomes, we obtain the probability for 5 channels mentioned in the main text: $p_1\approx0.31$ for doing nothing, $p_2\approx0.3$ for exchanging $e-m$, $p_3\approx0.12$ for measuring $f_x$, $p_4\approx0.23$ for measuring $f_z$, and $p_5\approx0.04$ for measuring $f_{xz}$. On the other hand, the space of all possible configurations that encode the two logical qubits is six-fold $\{(m_x,m_z),(e_x,e_z),(m_x,e_x),(m_z,e_z),(f_x,f_z),(m_{xz},e_{xz})\}$ (note that we only consider configurations with $x\leftrightarrow z$ symmetry given that the initial state and the measurement schedule are both symmetric). Depending on the initial configurations, not all transformations result in different outcomes, e.g. $(m_x,e_x)$ is invariant under $e-m$ exchange and $f_x$ measurement. Consequently, we obtain the transfer matrix in the basis of configurations previously mentioned
  \begin{equation}
  	S = \begin{pmatrix}
  	0.31 & 0.30  & 0 & 0 & 0 & 0 \\
  	0.30  & 0.31 & 0 & 0 & 0 & 0 \\
  	0.12 & 0.12 & 0.73 & 0 & 0 & 0\\
  	0.23 & 0.23 & 0 & 0.84 & 0 & 0 \\
  	0    &  0 &  0.27 & 0.16 & 1 & 0.35\\
  	0.04 & 0.04 & 0 & 0 &0 & 0.65
  	\end{pmatrix}
  \end{equation} 
  and $G(t)=\left(S^t\right)_{1,1}+\left(S^t\right)_{3,1}$. This expression with empirical $S$ agrees with the numerical data for all $t$ as shown in the main text. Matrix $S$ also yields the decay rate of $\approx0.30$ (corresponding to the logarithmic of the third largest eigenvalue) caused by the leaking to the steady configuration $(f_x,f_z)$. Away from critical, we know that the probability of measuring extended $f-$strings vanishes in the thermodynamic limit because the cluster boundaries are finite, leading to a zero decay rate. To justify this statement, we numerically extract the decay rate of the logical qubit over time by fitting $\tilde{G}(2t)=G(2t)+G(2t-1)$ to an exponential form $e^{-2\beta t}$. Fig.~\ref{Fig2}(b) shows that $\beta$ assumes an exponential decay with system size $\beta\sim e^{-L/\xi}$ away from the critical point but becomes $\beta\sim \mathcal{O}(L^0)$ corresponding to $\xi\to\infty$ at $p_M^c\approx 0.48$. The critical value of $\beta$ is consistent with those obtained from the percolation transition and from the scaling analysis of $G_0$ [see Fig.~\ref{Fig2}(a)]. The percolation picture holds quantitatively even if we modify the missing-link perturbation. In Fig.~\ref{Fig2}(c-d), we skip link measurements only in the green round instead of the green and blue rounds as in the main text. The FET-T transition is similar to the original model except for that $p_M^c\approx 0.65$, reflecting the percolation transition on a hexagonal lattice (instead of a kagome lattice). We note the percolation picture applies exactly as long as one color round is measured perfectly, making the ISG (except for the logical qubits) repeat itself after one cycle and the dynamics Markovian. In fact, if missing perturbations are introduced to all three color rounds, for sufficiently high $p_M$, randomness prevails over Floquet periodicity and drives the circuit into the critical phase that has super-area-law entanglement entropy, similar to that in the randomly measured Kitaev honeycomb model \cite{Lavasani2022,Sriram2022}.
	
\subsection{Role of the initial state}
\begin{figure}
    \centering
    \includegraphics[width=0.95\textwidth]{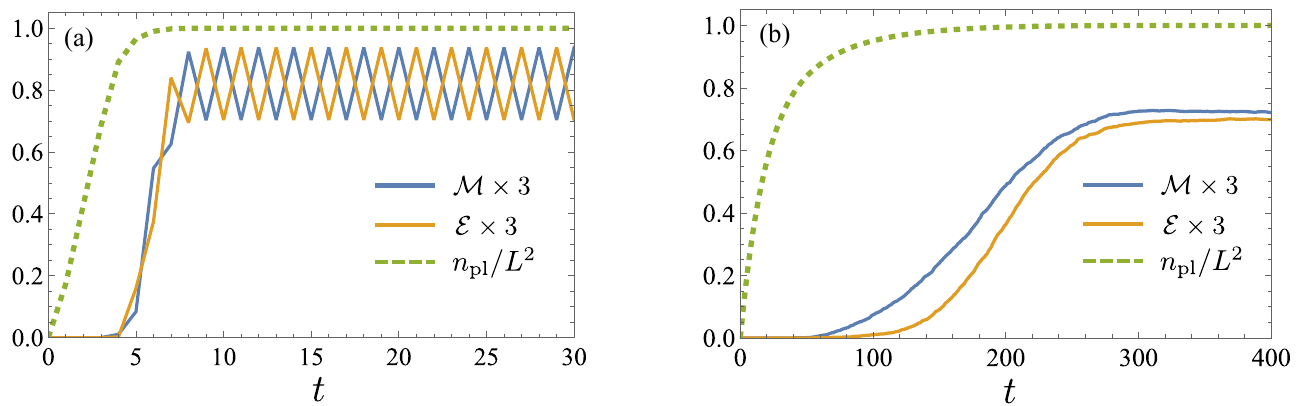}
    \caption{Evolution of plaquette occupation, $m$ readout (magnified 3 times), and $e$ readout (magnified 3 times) in the FET phase $p_M=0.3,~p_S=0$ (a) and stationary topological phase $p_M=0.6,~p_S=0$ (b). Note the different time scales between (a) and (b), as well as the rapid oscillations of $\mathcal{M}$ and $\mathcal{E}$ present in (a) but lacking in (b). The system linear size is 24, in each set of parameters, we randomly pick 5 initial states and perform 800 random realizations of the circuit on each initial state.}
    \label{initialstate}
\end{figure}

In the main text, we initialize the circuit by measuring all the plaquettes and two logical operators. We note that the Floquet-induced topological order, characterized by the $e-m$ exchange is a character of the measurement schedule and not the initial state. However, the initial state has an effect on the manifestation of FET. Specifically, if the state already has equal expectation values over $e$ and $m$ operators, the $e-m$ exchange of FET cannot be observed. In Fig.~\ref{initialstate}, we perform measurements on an ensemble of 5 random product states, i.e. each qubit can be independently oriented in the X, Y, or Z directions. For any value $p_M<1$, at a sufficiently late time, all plaquettes are measured, signaling the emergence of topological order. This time increases as $p_M\to 1$ but does not scale with the system size. The question is now whether this topological phase manifests time-crystalline order. By fluctuation of statistics, this rather small ensemble has unbalanced $\mathcal{M}$ and $\mathcal{E}$ readouts, where $\mathcal{M}(t) = \left(\overline{\langle{m_x(t)}\rangle^2} + \overline{\langle{m_z(t)}\rangle^2} \right)/2$ and $\mathcal{E}(t) = \left(\overline{\langle{e_x(t)}\rangle^2} + \overline{\langle{e_z(t)}\rangle^2} \right)/2$. Within the FET phase [Fig.~\ref{initialstate}(a)], this imbalance undergoes fast double-period oscillation; while in the stationary topological phase [Fig.~\ref{initialstate}(b)], we only observe a smooth and slow variation. This supports our claims that the phase diagram is independent of initialization. Our choice of initialization in the main text is for maximizing the manifestation of $e-m$ exchange by setting the initial state in the $(m_x, m_z)$ sector.

\subsection{Tripartite mutual information and purification dynamic}
\begin{figure}
	\centering
	\includegraphics[width=0.95\textwidth]{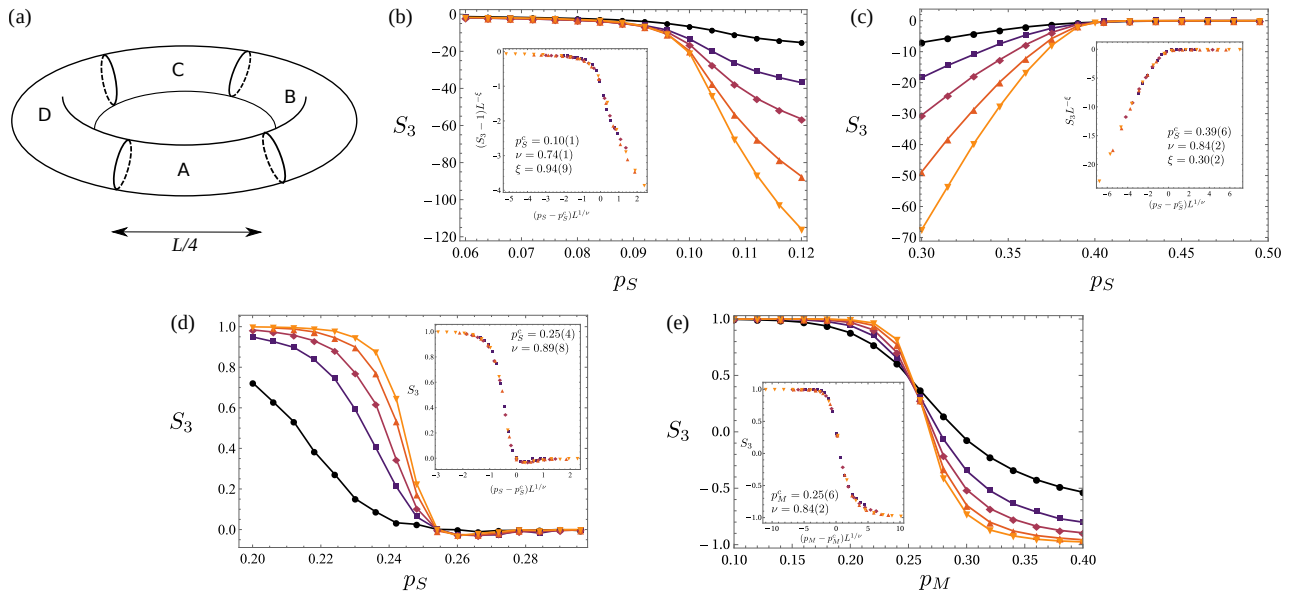}
	\caption{(a) Partition scheme to compute the tripartite mutual information. (b-d) TMI versus $p_S$ and data collapse for $p_M=0.4$ from the topological to volume-law phase (b), $p_M=0.45$ from the volume-law to confined area-law phase (c), and $p_M=0$ from the topological to confined area-law phase (d). (e) TMI versus $p_M$ and data collapse from the topological to the critical phase upon introducing missing perturbations to all color rounds. For (d) and (e), the scaling exponent is close to that of the 3D percolation transition, evident of the extension along time direction.}
	\label{Fig4}
\end{figure}

In this section, we provide the methodology and raw data for the tripartite mutual information (TMI). TMI is defined as 
\begin{equation}
	S_3 = S_A+S_B+S_C-S_{AB}-S_{BC}-S_{AC}+S_{ABC},
\end{equation} 	
where $S$ is the usual bipartite entanglement entropy and the partitions A, B, and C are illustrated in Fig.~\ref{Fig4}(a). We emphasize that this partition scheme is different from the topological entanglement entropy \cite{Kitaev2006} where A, B, and C share common edges pairwise. The reason to choose TMI over topological entanglement entropy is due to its capability of distinguishing the critical phase with super-area-law scaling. This phase does not exist in the main text but can be realized within our model given appropriate perturbation. In Fig.~\ref{Fig4}(b-d), we show the scaling analysis to obtain the critical point for the topological-volume law, volume law-area law, and topological-area law, respectively. Because TMI diverges in the volume-law phase, we use three scaling ansatzs $S_3=1+L^{\varepsilon}f[(p-p^c)L^{1/\nu}]$, $S_3=L^{\varepsilon}f[(p-p^c)L^{1/\nu}]$, and $S_3=f[(p-p^c)L^{1/\nu}]$ for the respective transition. 

In particular, the transition at $p_M=0$ has a scaling exponent $\nu\approx 0.89$ very close to that of the 3D percolation transition. Without missing errors, the action is one cycle is either measuring a plaquette through measuring all color links or measuring a single-qubit Pauli operator which anti-commutes with two adjacent plaquette operators. As such, the dynamics is similar to Ref.~\cite{Lavasani2020} albeit on a different geometry, where measuring a plaquette (single-qubit Pauli operator) is equivalent to adding (coalescing) colors of plaquettes. The trivial phase corresponds to a picture where a single color percolates throughout the entire bulk. This explain the 3D percolation universality of the transition. Out of the three dimensions, one is temporal; which is in contrast to the FET-T transition that belongs to the universality class of 2D percolation without the extension in time. We note that the 2D percolation universality is because red links are measured without defects, resetting the bulk stabilizers after every cycle. We can make the dynamics time-extensive by introducing missing perturbations to all three color rounds leading to the critical phase for large enough $p_M$. In Fig.~\ref{Fig4}(e), the TMI approaches $-1$ for $p_M > 0.26$, justifying the critical nature of the phase. This critical phase is similar to that of~\cite{Lavasani2022}. Here, open strings are allowed so we can keep track of the dynamics through the evolution of these strings, or more specifically their endpoints. For sufficiently large $p_M$, the moves of an endpoint are uncorrelated between far away time slices, making them essentially random walks and producing long strings at late time; while in the opposite limit, they are confined because of the mostly deterministic measurement schedule. The transition is also controlled by the 3D percolation universality class \cite{Lavasani2022}.

\begin{figure}
	\centering
	\includegraphics[width=0.95\textwidth]{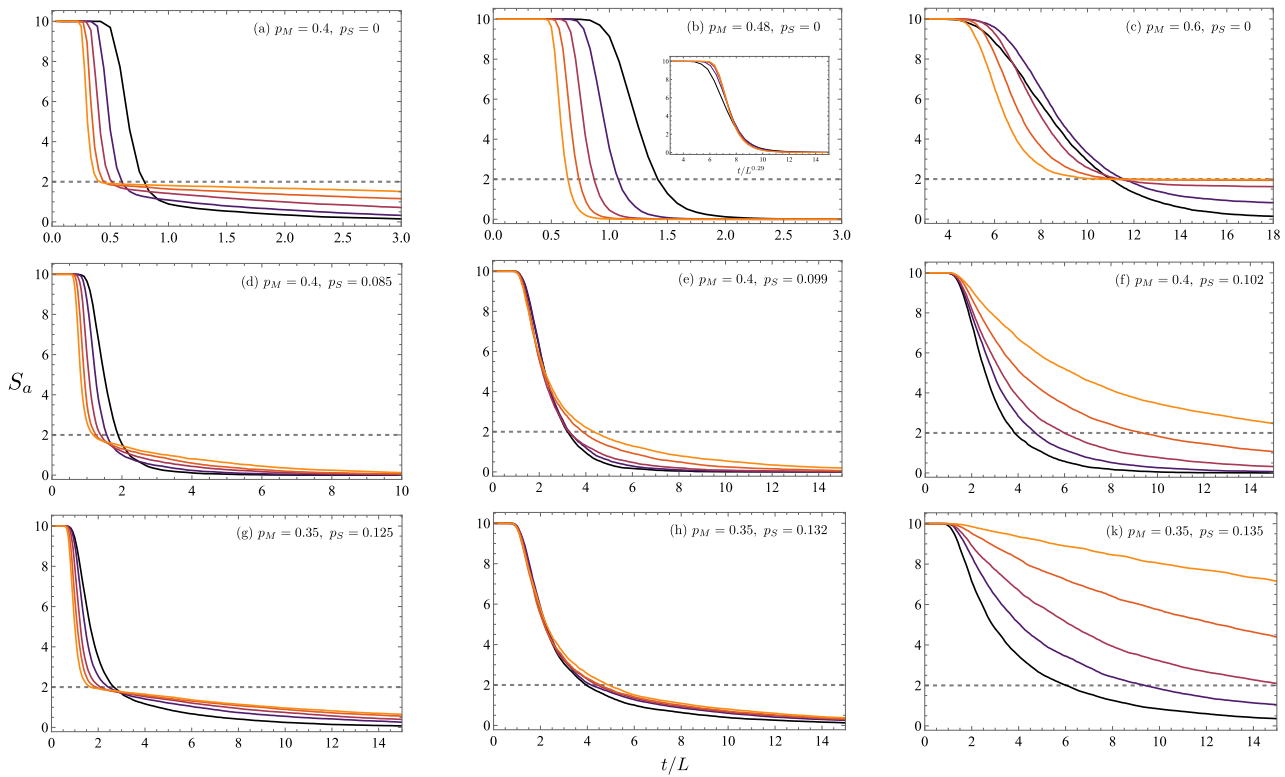}
	\caption{Entanglement entropy of the ancilla qubits. The parameter values of the middle column are tuned to produce the best-collapsed data with respect to $t/L$ except for the $p_S=0$ case. These values are to be compared with the critical value obtained from TMI: $(p_S,p_M)= (0,~0.477)$, $(0.101,~0.4)$ and $(0.130,~0.35)$. }
	\label{Fig5}
\end{figure}

We also check the consistency of the TMI with purification dynamics \cite{Gullans2020,Gullan2020b}. In this test, we initially scramble the honeycomb lattice plus 10 ancilla qubits by applying 4-qubit random Clifford gates over randomly chosen qubits (including the ancilla). The number of gates is $8L^3$. We then implement the measurement schedule described earlier and monitor the entanglement entropy of the ancilla qubits $S_a$ with respect to time. From the purification dynamics, we can estimate the location of the transition point, however, scaling analysis to exactly identify this point is not reliable given the small number of ancilla qubits. We expect that purification dynamics yields the identical transition to TMI if the number of ancilla qubits scales with that of the primary circuit. We first study the purification without $S-$perturbations ($p_S=0$) shown in Fig.~\ref{Fig5}(a-c). In both the FET and T phases, $S_a$ quickly reduces to 2, the number of logical qubits of the toric code, then decays at a rate exponentially slow with large $L$. On the contrary at the tricritical point, the ancilla qubits are purified within $t\sim L^{0.29}$. We note that the exponent of $0.29$ is slightly different from the $z=0$ in the main text probably because we do not initialize all the plaquettes at $t=0$ in the context of purification dynamics. Nevertheless, this value is still significantly lower than the usual $z=1$ expected for CFT at a critical point. Moving away from the tricritical point, we can see that the transition from topological to volume-law phase most likely has $z=1$ [see Fig.~\ref{Fig5}(e) and (h)]. The transition value estimation from ancilla purification is consistent with the value obtained from finite-size scaling of TMI.

 \section{Correct order parameter in the presence of single-qubit perturbation}
 Without $S-$perturbations, the stabilizer group in the bulk repeats itself \textbf{exactly} after one cycle, so we can probe the Floquet characteristic using the same loop operator that we initialize. In the presence of perturbations, the expectation value of a string operator is exponentially suppressed by its length because the loop may cut through an error and in fact this is guaranteed to happen as the length approaches infinity. This is similar to computing loop condensation is a dirty toric code even though the topological order is supposed to be robust against local perturbations. The solution is to morph the string to avoid all the errors. 
 
\begin{figure}
	\centering
	\includegraphics[width=0.85\textwidth]{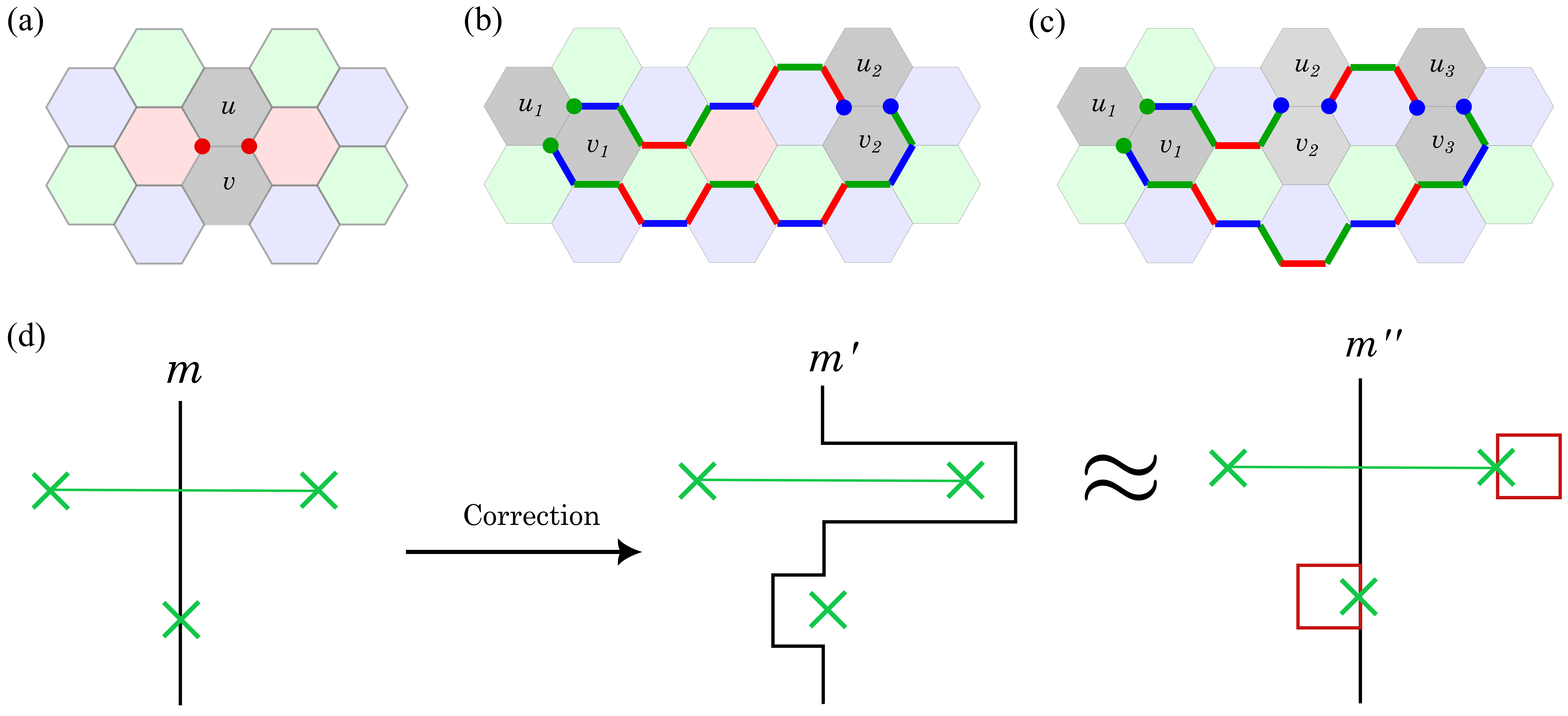}
	\caption{(a) Isolated single-qubit errors. (b) Errors located at the end points of an $f-$string. (c) Similar to (b) but three errors occur during the $f-$loop formation. (d) Adaption scheme involves deforming the $m-$string to avoid all the point and string errors. The action is equivalent to dressing the string with appropriate plaquettes.}
	\label{Fig6}
\end{figure}

 Without access to the many-body wavefunction of the circuit, the actual error configuration can be figured out by measuring the expectation value of plaquettes and their higher-order correlation. Specifically, an $S-$perturbation occurs along the shared link between the plaquettes $u$ and $v$, then we have the following identities $\mean{P_u}=\mean{P_v}=0$ and $\mean{P_uP_v}=\pm 1$. To identify if these perturbations exist by themselves or as end points of an $f-$string, we compute the long-range correlation function. For example, for an isolated $S-$perturbation depicted in Fig.~\ref{Fig6}(a), all the long-range correlation $\mean{P_u\Pi P_{k_i}}=0$ for all collections $\{k_i\} \not\subset\{u,v\}$. On the other hand, $f-$string errors can be detected through nontrivial correlation $\mean{P_{u_1}P_{u_2}}=\pm 1$ and $\mean{P_{u_1}P_{u_2}P_{u_3}} = \pm 1$, corresponding to Figs.~\ref{Fig6}(b) and (c) respectively. Once all the errors are mapped out, one can, in principle deform the $m-$string to avoid all these errors as shown in Fig.~\ref{Fig6}(d). We emphasize that in the topological phase, string errors must be finite, so we only need to examine errors within a finite strip around the $m-$string. 
 
\begin{figure}
	\centering
	\includegraphics[width=0.9\textwidth]{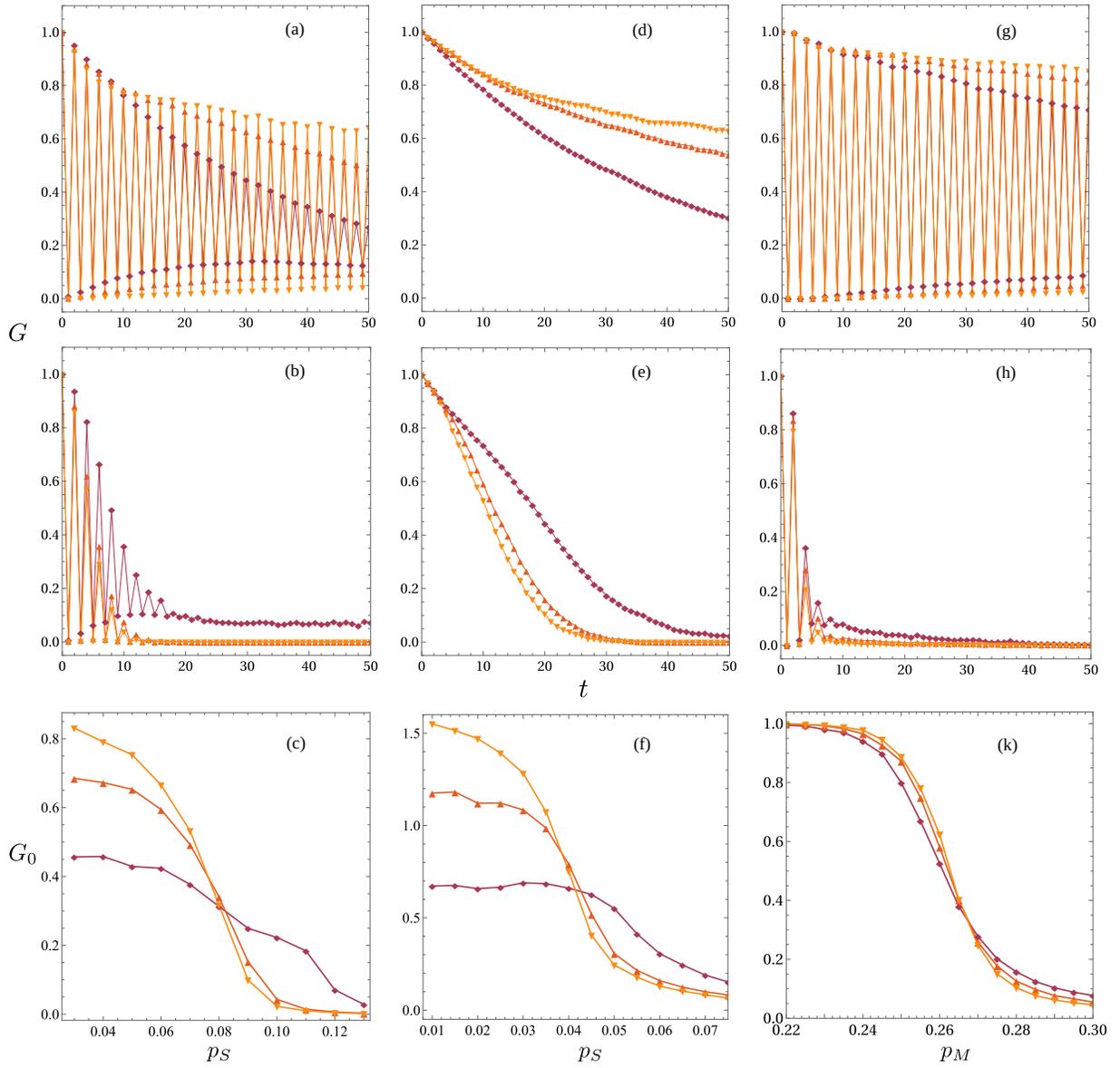}
	\caption{(a-c) $e-m$ exchange readout at $p_M=0.4$ in the FET phase ($p_S=0.06$) (a) and the volume-law phase ($p_S=0.11$) (b). (c) The zero component $G_0$ across the phase transition. (d-f) Similar to (a-c) but the transition from the stationary topological ($p_S=0.03$) (d) to the volume-law phase ($p_S=0.06$) at fixed $p_M=0.55$ (e). (g-k) Transition from the FET ($p_M=0.25$) (g) to the critical super-area-law phase ($p_M=0.3$) (h).}
	\label{Fig7}
\end{figure}

 In our work, by taking advantage of the full many-body wavefunction, we devise a different correction scheme that requires less overhead. The procedure we are going to describe below applies to a single random configuration. The result is then averaged over multiple configurations to get a representative picture of the phase. Intuitively, deforming the $m-$string is equivalent to dressing the string with appropriate plaquette operators that also anticommute with the same errors [see Fig.~\ref{Fig6}(d)]. The result is a composite object $m''=m\Pi_{u_i}P_{u_i}$ that commutes with all the errors. To identify whether such a $\{u_i\}$ collection exists, we first express the $m-$Pauli string as a $\mathbb{Z}_2^N$ vector with $N$ being the qubit number. Each entry $[m]_j=0$ if $m$ commutes with $j-$Pauli string of the stabilizer tableau and $1$ otherwise. If the expectation value $\mean{m}=\pm 1$, $m$ must commute with all rows of the stabilizer tableau, and thus $[m]=0$; while $\mean{m}=0$ implies that at least one entry is 1. We can perform the same transformation for all plaquette operators within a strip of width $d$ from the $m-$string and collect those that have non-trivial binary vector description as $P = \{[P_{u_1}],[P_{u_2}],...,[P_{u_n}]\}$. These plaquettes have zero expectation value and thus must overlap with an error. As mentioned earlier, some products among these plaquettes can have deterministic values, so $\text{rank}(P) < n$. If $\text{rank}(\{[m],P\}) = \text{rank}(P)$, then there exists a collection $\{i\}$ such that $[m'']=[m]\times \Pi [P_{u_i}]=0 $ or equivalently $\mean{m''}=\pm1$. Otherwise, we register a zero readout.  We apply the same procedure to study the $e-m$ exchange in the presence of single-qubit errors. The results are displayed in Fig.~\ref{Fig7}(a-f) with fixed $d=11$, showing that both the double and single-period behaviors characterizing the FET and T phases persist up to a finite $p_S$. In both cases, the critical value $p_S$ is somewhat less than the value obtained from TMI.
 
 In the topological-critical phase transition, when missing perturbations are present throughout the measurement schedule, the errors are not detectable by plaquette operators because they commute with all link measurements. However, the correction scheme we develop earlier can be modified by dressing the $m-$string with red link operators instead of hexagonal plaquettes. The red-dressed order parameter with fixed $d=7$ recovers the stable double-period oscillation in the Floquet topological phase [Fig.~\ref{Fig7}(g)], but fails to do so on the critical side [Fig.~\ref{Fig7}(h)]. From the oscillation of the corrected string readout, the phase transition happens around $p_M\approx 0.265$, which is surprisingly close to the value obtained from TMI, unlike the topological-volume law transition. We hypothesize that this difference is because the typical length of string errors diverges slower across the phase transition as the subsystem entanglement entropy $S\sim L\log L$ in the critical phase compared to $S\sim L^2$ in the volume-law phase.

\bibliographystyle{apsrev4-2}
\bibliography{reference}